\documentclass[aps,showpacs,amssymb,amsfonts,superscriptaddress,twocolumn,prl]{revtex4}
\usepackage{amsmath,bbm,mathrsfs}
\usepackage{graphicx}
\usepackage{amsfonts}
\usepackage{amssymb}
\usepackage{color}

\def\textbf#1{{\bf #1}}
\def\be{\begin{equation}}
\def\ee{\end{equation}}
\def\ben{\begin{eqnarray}}
\def\een{\end{eqnarray}}
\def\eea{\end{array}}
\def\bea{\begin{array}}
\newcommand{\ot}[0]{\otimes}
\newcommand{\Tr}[1]{\mathrm{Tr}#1}
\newcommand{\bei}{\begin{itemize}}
\newcommand{\eei}{\end{itemize}}
\newcommand{\ket}[1]{|#1\rangle}
\newcommand{\bra}[1]{\langle#1|}

\newcommand{\proj}[1]{\ket{#1}\!\bra{#1}}

\begin{document}

\newcommand{\eg}{{\it{e.g.~}}}
\newcommand{\ie}{{\it{i.e.~}}}
\newcommand{\etal}{{\it{et al.}}}
\newcommand{\daniel}[1]{{\color{red} #1}}
\newcommand{\remik}[1]{{\color{green} #1}}

\title{Perfect Quantum Privacy Implies Nonlocality}

\author{Remigiusz Augusiak}
\author{Daniel Cavalcanti}
\author{Giuseppe Prettico}
\affiliation{ICFO--Institut de Ci\`encies Fot\`oniques,
E--08860 Castelldefels, Barcelona, Spain}

\author{Antonio Ac\'in}
\affiliation{ICFO--Institut de Ci\`encies Fot\`oniques,
E--08860 Castelldefels, Barcelona, Spain}
\affiliation{ICREA--Instituci\'o Catalana de Recerca i Estudis
Avan\c{c}ats, Lluis Companys 23, 08010 Barcelona, Spain}

\begin{abstract}
Private states are those quantum states from which a perfectly
secure cryptographic key can be extracted. They represent the
basic unit of quantum privacy. In this work we show that all
states belonging to this class violate a Bell inequality. This
result establishes a connection between perfect privacy and
nonlocality in the quantum domain.
\end{abstract}


\maketitle

Classical and quantum information theory (QIT) are mainly theories
about resources~\cite{books}. Quantum features however make the
quantum theory richer and more powerful than its classical
counterpart. This richness is reflected by the variety of
different resources appearing in the quantum formalism. These are
for instance entanglement \cite{Hreview}, i.e., the existence of
compound states that do not admit a description in terms of
probabilistic combinations of products of states representing
individual subsystems, secret correlations \cite{crypto}, that is,
correlations that cannot be created by public communication, and
nonlocal correlations (see below) \cite{Scarani}. While some of
these resources, e.g., secret correlations, are also found in the
classical formalism, most of them do not have a classical
analogue. This allows performing tasks that are not achievable in
the classical world such as quantum teleportation
\cite{teleportation} or secure key distribution
\cite{BB84,Ekert91}. The two general questions the theory
addresses are (i) understanding those quantum resources necessary
to accomplish an information task and (ii) establishing
interconversion laws between all the different resources.

A key step when comparing and quantifying resources consists of
the identification of the basic unit for each of them. It is well
established that a Bell state, that is, a two-qubit maximally
entangled state, represents the basic unit of entanglement, known
as {\it e}-bit~\cite{e-bit}. Moving to secret correlations,
Horodecki {\it et al.} showed that private states are the basic
unit of privacy in the quantum domain~\cite{KH1,KH2}. Clearly, all
these states are entangled, as entanglement is a necessary
condition for secure key distribution~\cite{Curty,Acin-Gisin}.
However, a Bell state is just the simplest state belonging to the
larger class of private states. This implies that the distillation
of privacy from quantum states is not equivalent to entanglement
distillation, as it was commonly believed. Indeed, key
(entanglement) distillation from a quantum state $\rho$ can be
understood as the process of extracting copies of private (Bell)
states out of many copies of $\rho$. This nonequivalence is behind
the existence of bound entangled states that, though not allowing
for distillation of the Bell states \cite{BEnt98}, are a resource
for secure key distillation~\cite{KH1,KH2}.

Beyond these results, however, the principles allowing for secure
key distillation from quantum resources, a crucial question in
QIT, are hardly understood. In order to achieve this, it is
essential to identify the quantum properties common to all private
states. It is well known that Bell states are nonlocal since they
violate the Clauser-Horne-Shimony-Holt (CHSH) Bell inequality
\cite{CHSH}. Moved by this fact, one could ask whether all private
states violate a Bell inequality. This is {\it a priori} unclear,
as private states may exhibit radically different entanglement
properties~\cite{KH2}.

In this work we address the above question and show that all
private states are indeed nonlocal. This result is general, as our
proof works for any dimension and any number of parties. Private
states, then, not only represent the unit of quantum privacy, but
also allow two distant parties to establish a different quantum
resource, namely, nonlocal correlations. These states contain the
strongest form of entanglement as they can give rise to
correlations with no classical analogue. More generally, our
findings point out an intriguing connection between two of the
most intrinsic quantum properties: privacy and nonlocality.

{\it Preliminaries.--}Before proceeding with the proof of our
results, we recall in what follows the notions of nonlocality and
private states.

Consider first a Bell-type experiment in which party $i$ can
measure one of the $k_{i}$ observables $\{A_{i}^{(j_i)}\}$
$(j_i=1,\ldots,k_i)$, each with $r_{i}^{(j_i)}$ outcomes denoted
by $a_{i}^{(j_i)} \in \{1,\ldots,r_{i}^{(j_i)}\}$. We say that
there exists a local model for this experiment if the conditional
probabilities
$P(a_{1}^{(j_1)},\ldots,a_{N}^{(j_N)}|A_{1}^{(j_1)},\ldots,A_{N}^{(j_N)})$
of obtaining result $a_{i}^{(j_i)}$ upon the measurement of
$A_{i}^{(j_i)}$, can be written in the following form
\begin{eqnarray}\label{local}
&&P(a_{1}^{(j_1)},\ldots,
a_{N}^{(j_N)}|A_{1}^{(j_1)},\ldots,A_{N}^{(j_N)})=\\
&&\int\mathrm{d}\lambda
P(\lambda)P(a_{1}^{(j_1)}|A_{1}^{(j_1)},\lambda)\cdot \ldots\cdot
P(a_{N}^{(j_{N})}|A_{N}^{(j_N)},\lambda).\nonumber
\end{eqnarray}
Fine \cite{Fine} showed that the existence of this model for the
experiment is equivalent to the existence of a joint probability
distribution $P(a_{1}^{(1)},\ldots,a_1^{(k_1)},
\ldots,a_{N}^{(1)},\ldots,a_N^{(k_N)})$ involving all local
measurements, such that the marginal probabilities reproduce the
observed measured outcomes. The observed correlations are said to
be nonlocal if the conditional probability distributions
$P(a_{1}^{(j_1)},\ldots,a_{N}^{(j_N)}|A_{1}^{(j_1)},\ldots,A_{N}^{(j_N)})$
do not admit a local model. An $N$-partite quantum state $\rho_N$
is then nonlocal whenever it is possible to find local
measurements leading to nonlocal correlations when applied to
$\rho_N$.

Now, let us pass to the definition of private states
\cite{KH1,KH2,PHRA,RAPH}. In general, these are $N$--partite
states that can be written as
\begin{equation}\label{private state}
\Gamma_{\mathsf{AA}'}^{(d)}=\frac{1}{d}\sum_{i,j=0}^{d-1}(\ket{i}\!\bra{j})^{\ot
N}_{\mathsf{A}} \ot U_{i}\rho_{\mathsf{A}'}U_{j}^{\dagger},
\end{equation}
where $\rho_{\mathsf{A}'}$ is some density matrix, $\{U_{i}\}$  a
set of unitary operations, and $\mathsf{A}=A_{1}\ldots A_{N}$ and
$\mathsf{A}'=A_{1}'\ldots A_{N}'$ are multi--indices referring to
subsystems. The subsystem marked with the subscript $\mathsf{A}$
consists of $N$ qudits and is called the key part. The remaining
subsystem is the shield part and is defined on some arbitrary
finite-dimensional product Hilbert space
$\mathcal{H}'=\mathcal{H}_{1}'\ot\ldots\ot\mathcal{H}_{N}'$. Party
$i$ holds one particle from the key part $A_{i}$ and one from the
shield part $A_{i}'$. The key point behind the private states is
that $\log_2 d$ bits of perfectly secure bits of cryptographic key
can be extracted from
$\Gamma_{\mathsf{AA'}}^{(d)}$~\cite{KH1,Pankowski}.

{\it All private states are nonlocal.--}We are in position to
prove our main result. We divide the proof into two parts. First,
following the ideas of Ref.~\cite{PHRA}, we show that using local
quantum operations (represented by appropriately chosen quantum
channels) {\it without} any use of classical communication, the
key part of any private state (subsystem $\mathsf{A}$), can be
brought to the form
\begin{equation}\label{state}
\varrho_{N}^{(d)}=\sum_{k,l=0}^{d-1}\alpha_{kl}(\ket{k}\!\bra{l})^{\ot
N}
\end{equation}
with $\alpha_{kk}=1/d$ and at least one off-diagonal element
nonzero; i.e., there exists a pair of indices $k<l$ such that
$\alpha_{kl}\neq 0$. Note that the shield part is discarded during
this process. Second, we show that any state of the
form~\eqref{state} with $\alpha_{kl}\neq 0$ is nonlocal. Finally,
the fact that local operations without classical communication
cannot produce a nonlocal state from a local one implies that all
private states are nonlocal.


Let us now proceed with the first part of the proof. For this aim
we assume that the $i$th party performs, on its subsystems $A_{i}$
and $A_{i}'$, the quantum operation represented by the following
quantum channel
\begin{equation}\label{locop}
\Lambda^{(i)}(\cdot)=V^{(i)}(\cdot)V^{(i)\dagger}
+W^{(i)}(\cdot)W^{(i)\dagger},
\end{equation}
where the Kraus operators $V^{(i)}$ and $W^{(i)}$ are given by
\begin{equation*}
V^{(i)}=\sum_{k}\proj{k}_{A_{i}}\ot \widetilde{V}_{k}^{(i)},\qquad
W^{(i)}=\sum_{k}\proj{k}_{A_{i}}\ot \widetilde{W}_{k}^{(i)},
\end{equation*}
The operators $\widetilde{V}_{k}^{(i)}$ and
$\widetilde{W}_{k}^{(i)}$ act on the shield part belonging to the
$i$th party (the $A_{i}'$ subsystem) and are chosen so that they
define a proper quantum measurement. Precisely, given
$\widetilde{V}_{k}^{(i)}$ we define the second Kraus operator to
be
$\widetilde{W}_{k}^{(i)}=(\mathbbm{1}-\widetilde{V}_{k}^{(i)\dagger}\widetilde{V}_{k}^{(i)})^{1/2}$,
with $\mathbbm{1}$ being the identity matrix acting on the
$A_{i}'$ subsystem. Application of all the channels
$\Lambda^{(i)}$ to $\Gamma_{\mathsf{AA}'}^{(d)}$ results in the
following state
\begin{equation*}
\bigotimes_{i=1}^{N}\Lambda^{(i)}(\Gamma_{\mathsf{AA}'}^{(d)})=\frac{1}{d}\sum_{k,l=0}^{d-1}\ket{k}\!\bra{l}^{\ot
N}\ot
\sum_{n=1}^{2^{N}}X_{k}^{(n)}U_{k}\varrho_{\mathsf{A}'}U_{l}^{\dagger}X_{l}^{(n)\dagger},
\end{equation*}
where matrices $X_{k}^{(n)}$ are defined as members of the
$2^N$--element set
$\{\widetilde{V}_{k}^{(i)},\widetilde{W}_{k}^{(i)}\}^{\ot N}$.
Explicitly, one has
$X_{k}^{(1)}=\widetilde{V}_{k}^{(1)}\ot\ldots\ot
\widetilde{V}_{k}^{(N)},
X_{k}^{(2)}=\widetilde{V}_{k}^{(1)}\ot\ldots\ot\widetilde{V}_{k}^{(N-1)}\ot
\widetilde{W}_{k}^{(N)}$, and so on. Tracing now the shield part
we get the promised state (\ref{state}) with $\alpha_{kl}$ given
by
\begin{equation}\label{application}
\alpha_{kl}=\Tr\left[\bigotimes_{i=1}^{N}\left(\widetilde{V}_{l}^{(i)\dagger}\widetilde{V}_{k}^{(i)}
+\widetilde{W}_{l}^{(i)\dagger}\widetilde{W}_{k}^{(i)}\right)U_{k}\varrho U_{l}^{\dagger}\right].
\end{equation}
One also finds that, since by construction
$\widetilde{V}_{k}^{(i)\dagger}\widetilde{V}_{k}^{(i)}
+\widetilde{W}_{k}^{(i)\dagger}\widetilde{W}_{k}^{(i)}=\mathbbm{1}$
for any $i$, the diagonal elements $ \alpha_{kk}$ of this state
are equal to $1/d$.

Now we need to show that at least one of the above coefficients is
nonzero. In other words, for some fixed pair of $k$ and $l$
$(k<l)$ we need to choose the operators $\widetilde{V}_{k}^{(i)}$
and $\widetilde{V}_{l}^{(i)}$ in such a way that $\alpha_{kl}$ is
nonzero. To this aim we simplify a little our considerations by
assuming that the operators $\widetilde{V}_{k}^{(i)}$ and
$\widetilde{V}_{l}^{(i)}$ corresponding to $i$th party are
positive and diagonal in the same basis. Thus, we can write these
particular operators in the form
\begin{equation*}\label{Vdiag1}
\widetilde{V}_{k}^{(i)}=\sum_{m}v_{m}^{(i)}\proj{e_{m}^{(i)}},\qquad
\widetilde{V}_{l}^{(i)}=\sum_{m}\overline{v}_{m}^{(i)}\proj{e_{m}^{(i)}},
\end{equation*}
where we assume that the eigenvalues satisfy
$v_{m}^{(i)},\overline{v}_{m}^{(i)}\in [0,1]$ and the eigenvectors
$\ket{e_{m}^{(i)}}$  are orthonormal, i.e., $\langle
e_{m}^{(i)}|e_{n}^{(i)}\rangle=\delta_{mn}$ (note that the fixed
indices $k,l$ we are interested in are omitted in the right--hand
side of the previous expression). This, in turn means that the
operators $\widetilde{W}_{k}^{(i)}$ and $\widetilde{W}_{l}^{(i)}$
are also diagonal in the basis $\{\ket{e_{m}^{(i)}}\}$, and have
eigenvalues $(1-v_{m}^{(i)2})^{1/2}$ and
$(1-\overline{v}_{m}^{(i)2})^{1/2}$, respectively. As a
consequence the operator appearing in parenthesis in Eq.
\eqref{application} simplifies to
\begin{equation}\label{operatorDiag}
\widetilde{V}_{l}^{(i)\dagger}\widetilde{V}_{k}^{(i)}
+\widetilde{W}_{l}^{(i)\dagger}\widetilde{W}_{k}^{(i)}=\sum_{m}\beta_{m}^{(i)}\proj{e_{m}^{(i)}},
\end{equation}
where its eigenvalues are given by
$\beta_{m}^{(i)}=v_{m}^{(i)}\overline{v}_{m}^{(i)}+(1-v_{m}^{(i)2})^{1/2}(1-\overline{v}_{m}^{(i)2})^{1/2}$
and obviously satisfy $0\leq \beta_{m}^{(i)}\leq 1$. Now, putting
Eq. (\ref{operatorDiag}) to Eq. (\ref{application}), we get
\begin{eqnarray}\label{alphas}
\alpha_{kl}&=&\sum_{m_{1},\ldots,m_{N}}\beta_{m_{1}}^{(1)}\ldots\beta_{m_{N}}^{(N)}\nonumber\\
&&\times\bra{e_{m_{1}}^{(1)}}\ldots \bra{e_{m_{N}}^{(N)}}U_{k}\rho
U_{l}^{\dagger}\ket{e_{m_{1}}^{(1)}}\ldots \ket{e_{m_{N}}^{(N)}}.
\end{eqnarray}
Finally, to prove that $\alpha_{kl}\neq 0$ it suffices to notice
that for any nonzero matrix $X$ (and in particular $U_{k}\rho
U_{l}^{\dagger}$) there always exists at least one $N$ partite
product vector $\ket{\psi}=\ket{\psi_{1}}\ldots\ket{\psi_{N}}$
such that $\langle\psi| X|\psi\rangle$ is nonzero. Otherwise, if
for all such vectors $\langle\psi| X|\psi\rangle=0$, the matrix
$X$ has to be the zero matrix (see Lemma 2 of
Ref.~\cite{Zyczkowski}).

As just discussed, there exists a product vector $\ket{\psi}$ such
that $\langle\psi|U_{k}\rho U_{l}^{\dagger}|\psi\rangle\neq 0$ for
a pair of indices $k<l$. Therefore we can always chose
$\widetilde{V}_{k}^{(i)}$ and $\widetilde{V}_{l}^{(i)}$ for each
party in such way that $\ket{\psi}$ is one of the product vectors
appearing in Eq. (\ref{alphas}) (more precisely, $\ket{\psi_{i}}$
can be set as one of eigenvectors of $\widetilde{V}_{k}^{(i)}$ and
$\widetilde{V}_{l}^{(i)}$). Now, we can use the freedom in the
numbers $\beta_{m}^{(i)}$ in such a way that $\alpha_{kl}\neq 0$,
which is exactly what we wanted to prove. Actually, we can always
choose $\widetilde{V}^{(i)}_{k(l)}$ so that at least one of the
coefficients $\alpha$'s in each row and column of
$\varrho_{N}^{(d)}$ is nonzero.


Let us move to the second part of the proof. In what follows we
show that any state of the form (\ref{state}) is nonlocal. First
we will consider the bipartite case and then we will move to the
multipartite scenario.

{\it Bipartite case} ($d=2$).--A generic form of the simplest
example of bipartite private states (two--qubit key part) reads
(zeros denote null matrices of adequate dimension)
\begin{equation}\label{2qubitTens}
\Gamma_{\mathsf{AA'}}^{(2)}=\frac{1}{2}
\left(\begin{array}{cccc}U_{0}\rho_{\mathsf{A}'} U^{\dagger} _{0}
& 0 & 0 & U_{0}\rho_{\mathsf{A}'}U^{\dagger} _{1}
\\ 0 & 0 & 0 & 0 \\ 0 & 0 & 0 & 0 \\ U_{1}\rho_{\mathsf{A}'}
U^{\dagger} _{0} & 0 & 0 & U_{1}\rho_{\mathsf{A}'}U^{\dagger}
_{1}\end{array}\right) .
\end{equation}
After applying the previous local quantum operations to this state
the parties are left with a two-qubit state:
\begin{equation}\label{2qubit}
\varrho_{2}^{(2)}=
\left(\begin{array}{cccc}1/2 & 0 & 0 & \alpha_{01} \\
0 & 0 & 0 & 0 \\ 0 & 0 & 0 & 0 \\
\alpha_{01}^{*} & 0 & 0 & 1/2\end{array}\right).
\end{equation}
Since we already know that $\alpha_{01}\neq 0$, it follows from
the criterion proposed in Ref.~\cite{Horodeccy} that the above
state violates the CHSH-Bell inequality~\cite{CHSH} (here written
in the equivalent Clauser-Horne form~\cite{CH})
\begin{eqnarray}\label{CH}
P(A_1 B_1)+P(A_2 B_1)&+&P(A_1 B_2) -P(A_2 B_2)\nonumber\\ &-&
P(A_1)-P(B_1)\leq 0.
\end{eqnarray}
Here $P(A_iB_j)$ denotes the probability that Alice and Bob obtain
the first result upon the measurement of observables $A_i$ and
$B_j$ $(i,j=1,2)$.
Recall that the CHSH test involves the measurement of two
dichotomic observables per site.

{\it Bipartite case} ($d>2$).--For higher dimensional bipartite
private states we use the fact that the inequality \eqref{CH} only
involves one measurement outcome for each of the observables. For
this purpose, let us first assume that some $\alpha_{kl}$ is
nonzero and rewrite $\varrho_{2}^{(d)}$ (cf. Eq. (\ref{state})) as
\begin{eqnarray}\label{dit}
\varrho_{2}^{(d)} = \left(
\begin{array}{ccccc}
\ddots & \vdots & & \vdots & \\
\cdots & 1/d & \cdots & \alpha_{kl} & \cdots  \\
& \vdots & \ddots & \vdots &         \\
\cdots & \alpha_{kl}^{*} & \cdots & 1/d & \cdots  \\
& \vdots &        & \vdots & \ddots  \\
\end{array}
\right).
\end{eqnarray}
The marked $2\times 2$ submatrix can be seen, up to a
normalization factor $2/d$, as a two-qubit state like the one
given in Eq. \eqref{2qubit}. As we have just shown, any such
two-qubit state with nonzero off-diagonal element is nonlocal.
Therefore, to prove nonlocality of $\varrho_{2}^{(d)}$ we can
design the observables $A_{i}$ and $B_{i}$ $(i=1,2)$ so that their
first outcomes correspond to one-qubit projectors (embedded in
$\mathbb{C}^{d}$) leading to the violation of (\ref{CH}) by the
corresponding two-qubit state. Precisely, we take the projectors
$\mathcal{P}_{A}^{(i)}=\proj{\psi_{i}}$ and
$\mathcal{P}_{B}^{(i)}=\proj{\widetilde{\psi}_{i}}$ $(i=1,2)$,
where the pure states $\ket{\psi_{i}}$ and
$\ket{\widetilde{\psi}_{i}}$ are of the general one-qubit form
$a\ket{k}+b\ket{l}$. The remaining outcomes (which are irrelevant
from the point of view of the inequality (\ref{CH})) of the
involved observables $A_i (B_i)$ can just correspond to projectors
$\mathbbm{1}-\mathcal{P}^{(i)}_{A(B)}$ $(i=1,2)$.

Now, by using these settings in the CHSH test \eqref{CH}, one sees
that the state \eqref{dit} leads to almost the same violation as
for the two-qubit state in Eq.~\eqref{2qubit} with the only
difference being the normalization factor $2/d$. Clearly, this
does not cause any problem since the same factor appears in all
the terms of the inequality. Therefore it {\it does not} change
the sign of the CHSH parameter~\eqref{CH}. As a conclusion the
CHSH-Bell inequality for any bipartite state $\varrho_{2}^{(d)}$
is also violated.

{\it Multipartite case.--}We now move to the multipartite case. In
order to prove the nonlocality of the states \eqref{state} we
exploit the fact that, given a generic $N$-partite state,
$\rho_N$, if there exist local projections of $N-m$ particles onto
a product state leaving the remaining $m$ particles in a nonlocal
state, $\rho_m$, the initial state $\rho_N$ is nonlocal. This
follows from the fact that one cannot produce in this way a
nonlocal state from a local one. The same reasoning was used,
e.g., in Ref.~\cite{PopescuRohrlich} in the context of proving the
nonlocality of general multipartite pure entangled states.

Indeed, denote by $A_i$ $(i=m+1,\ldots,N)$ the local measurements
(with outcomes $a_i$) by the previous $N-m$ parties such that for
one of the outcomes, say 0, the state $\rho_m$ shared by the
remaining $m$ parties is nonlocal. For the sake of simplicity we
assume that the nonlocality of this $m$-partite state can be
proven with only two measurements per site, $A_i$ and $A'_i$ with
outcomes $a_i$ and $a'_i$ $(i=1,\ldots,m)$ (our reasoning can be
trivially adapted to Bell tests involving more measurements).
According to Fine's result (see above), there cannot exist a joint
probability distribution
$P(a_1,a'_1,\ldots,a_m,a'_m|a_{m+1}=0,\ldots,a_N=0)$ reproducing
the observed outcomes for the $m$ parties conditioned on the fact
that the measurement result for the remaining $N-m$ parties was
equal to 0. Now, consider a Bell test for the initial $N$-partite
state $\rho_N$ where the parties apply all the previously
introduced measurements. Assume that the obtained statistics can
be described by a local model. Then, there exists a joint
probability distribution
$P(a_1,a'_1,\ldots,a_m,a'_m,a_{m+1},\ldots,a_N)$. But this would
immediately imply the existence of the joint probability
distribution $P(a_1,a'_1,\ldots,a_m,a'_m|a_{m+1}=0,\ldots,a_N=0)$,
which is in contradiction with the fact that $\varrho_m$ is
nonlocal. Thus, the initial state $\rho_N$ has to be nonlocal.

Using this argument, in order to prove the nonlocality of
multipartite states $\varrho_{N}^{(d)}$ it is enough to build
local projections mapping these states into a nonlocal state of a
fewer number of particles. Consider the local projections
$P_{\phi}$ onto
$\ket{\phi}=(1/\sqrt{d})(\ket{0}+\ldots+\ket{d-1})$. Projecting an
arbitrary subset of $N-m$ particles of $\varrho_{N}^{(d)}$ onto
$P_{\phi}$ the remaining $m$ parties are left with following
$m$--partite state
\begin{equation}\label{state2}
\varrho^{(d)}_{m}=\sum_{k,l=0}^{d-1}\alpha_{kl}(\ket{k}\!\bra{l})^{\ot
m}.
\end{equation}
Thus, if $N-2$ parties apply the projector $P_{\phi}$ to the state
\eqref{state}, the remaining two parties are left with a bipartite
private state $\varrho^{(d)}_{2}$. However, we have just shown
that this state is nonlocal. Thus, $\varrho_{N}^{(d)}$ must also be nonlocal.

{\it Discussion.--}Private states play a relevant role in QIT
because they represent perfectly secure bits of cryptographic
key~\cite{KH1,KH2,RAPH}. Knowing their entanglement properties is
crucial to understand the mechanism allowing for secure key
distribution from quantum states. In general, private states are
thought to have a weaker form of entanglement than Bell states.
However, we have shown here that all private states are nonlocal.
They have, then, the strongest form of quantum correlations, since
the results of local measurements on these states cannot be
reproduced by classical means.

Finally, it would be interesting to study how our findings can be
related to the Peres conjecture~\cite{PeresConj}, a long-standing
open question in quantum information theory. This conjecture
states that bound entangled states do not violate any Bell
inequality. The intuition is that these states have a very weak
form of quantum correlations. Then, all the correlations obtained
from these states should have a classical description. Note,
however, that there exist bound entangled states with positive
partial transposition which are arbitrarily close (in the trace
norm) to private states~\cite{KH1,KH2,Pankowski,RAPH,RAPH2}. This
is indeed the reason why these examples of bound entangled states
have nonzero distillable cryptographic key. But, as shown here,
all private states are nonlocal. One would then be tempted to
conclude that these bound entangled states are also nonlocal.
Interestingly, the situation is subtler than initially thought. In
fact, recall that the nonlocality of private states has been
proven here by showing the violation of the CHSH-Bell inequality.
Unfortunately, this inequality cannot be violated by bound
entangled states with positive partial transposition~\cite{WW}.
This implies that the violation of this inequality by private
states arbitrarily close to bound entangled states has to be very
small. In view of all these findings it appears interesting to
analyze the nonlocal properties of bound entangled states with
positive distillable secret key.

\begin{acknowledgments}We thank P. Horodecki, M. Lewenstein, and J. Stasi\'nska
for useful comments. This work was financially supported by the EU
Integrated Projects SCALA, AQUTE, and QAP, the ERC Starting Grant
PERCENT, the EU STREP NAMEQUAM and COMPAS, the Spanish MEC
projects QTIT (FIS2007-60182) and TOQATA (FIS2008-00784) and
Consolider-Ingenio QOIT projects, the Generalitat de Catalunya and
Caixa Manresa.
\end{acknowledgments}


\begin{thebibliography}{0}



\bibitem{books} T. M. Cover and J. A. Thomas, {\it Elements of Information Theory}
(Wiley, New York, 1991); M. A. Nielsen and I. L. Chuang, {\it
Quantum Computation and Quantum Information} (Cambridge University
Press, Cambridge, 2000).


\bibitem{Hreview}R. Horodecki \etal, Rev. Mod. Phys. {\bf 81}, 865
(2009).

\bibitem{crypto}N. Gisin \etal., Rev. Mod. Phys. {\bf 74}, 145
(2002).

\bibitem{Scarani}V. Scarani, {\it Quantum information: primitive notions
and quantum correlations}, arXiv:0910.4222.


\bibitem{teleportation}C. H. Bennett \etal., Phys. Rev. Lett. {\bf 70},
1895 (1993).

\bibitem{BB84}C. H. Bennett and G. Brassard,
{\it Proceedings IEEE Int. Conf. on Computers, Systems and Signal
Processing, Bangalore, India} (IEEE, New York, 1984), p. 175.

\bibitem{Ekert91}A. K. Ekert, Phys. Rev. Lett. {\bf 67}, 661 (1991).

\bibitem{e-bit}C. H. Bennett {\it et al.}, Phys. Rev. A {\bf 53}, 2046 (1996).


\bibitem{KH1}K. Horodecki {\it et al.}, Phys. Rev. Lett. {\bf 94}, 160502 (2005).


\bibitem{KH2}K. Horodecki {\it et al.}, IEEE Trans. Inf. Theory {\bf 55}, 1898 (2009).


\bibitem{Curty}M. Curty {\it et al.},
Phys. Rev. Lett. {\bf 92},
217903 (2004); M. Curty {\it et al.}, Phys. Rev. A {\bf 71},
022306 (2005).


\bibitem{Acin-Gisin}A. Ac\'in and N. Gisin, Phys. Rev. Lett. {\bf 94}, 020501 (2005).

\bibitem{BEnt98}M. Horodecki, P. Horodecki, and R. Horodecki, Phys. Rev. Lett. {\bf 80},
5239 (1998).

\bibitem{CHSH}J. F. Clauser {\it et al.}, Phys. Rev. Lett. {\bf 24}, 549 (1970).

\bibitem{Fine}A. Fine, Phys. Rev. Lett. {\bf 48}, 291 (1982).


\bibitem{PHRA}P. Horodecki and R. Augusiak, Phys. Rev. A {\bf 74}, 010302(R) (2006).


\bibitem{RAPH}R. Augusiak and P. Horodecki, Phys. Rev. A {\bf 80}, 042307 (2009).

\bibitem{Pankowski}K. Horodecki {\it et al.}, IEEE Trans. Inf. Theory {\bf 54}, 2621 (2008).

\bibitem{Zyczkowski}K. \.Zyczkowski {\it et al.}, Phys. Rev. A {\bf 58}, 883 (1998).

\bibitem{Horodeccy}R. Horodecki {\it et al.},
Phys. Lett. A {\bf 200}, 340 (1995).


\bibitem{CH}J. F. Clauser and M. A. Horne, Phys. Rev. D {\bf 10}, 526 (1974).

\bibitem{PopescuRohrlich}S. Popescu and D. Rohrlich, Phys. Lett. A {\bf 166}, 293 (1992).

\bibitem{PeresConj}A. Peres, Found. Phys. {\bf 29}, 589 (1999).

\bibitem{RAPH2}R. Augusiak and P. Horodecki, EPL {\bf 85}, 50001 (2009).

\bibitem{WW}R. F. Werner and M. M. Wolf, Phys. Rev. A {\bf 61}, 062102 (2000).



\end{thebibliography}
\end{document}